# Towards a Mechanical Analogy of a Quantum Particle: Turbulent Advection of a Fluid Discontinuity and Schroedinger Mechanics.

**Valery P. Dmitriyev**

*Lomonosov University, Moscow, Russia*
*P.O. Box 160, Moscow 117574, Russia*
*e-mail: dmitr@cc.nifhi.ac.ru*

## Abstract

A discontinuity of a turbulent ideal fluid is considered. It is supposed to be split and dispersed, or spread in the stochastic environment forming a gas without hydrostatic pressure. Two equal-mass fragments of a discontinuity are indistinguishable from each other. A gas, that possesses such properties, must behave itself as the Madelung medium.

## 1  Searching for Madelung medium

We are in search of a realization for mechanical medium, which behaves itself as the Madelung gas [1]. A discontinuity – void or phase precipitate – of a turbulent inviscid incompressible fluid is considered. It is split and spread in the stochastic environment forming the dispersion or plasma of the discontinuity. The latter can be viewed as a gas, with splinters of the discontinuity as its constitutive elements. This is rather an unusual gas. Splinters may penetrate freely through each other. Two equi-"mass" splinters are indistinguishable and hence interchangeable with each other. A splinter can easily grow or diminish in size at the expense of other parts of the discontinuity. Moreover, if even not changing in its parameters, all the same, the gas element loses its authenticity (self-equality) in the course of evolution. Therefore, unlike a particle of a material medium, it has no trajectory of motion in principle. Those properties are favorable in order to try the plasma of a point discontinuity as a mechanical model of a quantum particle.

## 2   Active and passive media

Dispersion of a void in a turbulent fluid represents a kind of solution. In some respects, it behaves itself like a drop of an admixture dissolved in a fluid. In a dilute solution, a solute is driven by a solvent. So, it is a passive scalar and the solvent is an active medium [2]. They say that a passive scalar is advected by an active medium. In general, we consider inhomogeneity (defect, imperfection or disturbance) of an active medium. It is supposed to be dispersed or distributed over the medium forming a scalar field $r(\mathbf{x},t)$ of the defect's density. It may be an impurity dissolved in a fluid, dislocation in a metal, a void in a fluid, heat in a medium, an eddy in a fluid etc.

One can't describe evolution of the density $V(\mathbf{x},t)$ of an active medium not examining simultaneously evolution of its velocity field $\mathbf{u}(\mathbf{x},t)$. In this event, two equations are needed – the mass balance

$$\partial_t V + \nabla \cdot (V\mathbf{u}) = 0 \tag{2.1}$$

and momentum balance

$$\frac{d\mathbf{u}}{dt} = \mathbf{f}. \tag{2.2}$$

In hydrodynamics

$$\mathbf{f} = -\nabla p.$$

For incompressible fluid (2.1) degenerates to

$$\nabla \cdot \mathbf{u} = 0. \tag{2.3}$$

The theory of a passive scalar features explicitly only kinematics of scalar's motion, all the dynamics being relegated to an active medium. We consider the scalars, which conserve an extensive characteristic (mass) in the course of advection. This is volume for voids, energy for heat and eddies, Burgers vector for dislocations etc. Evolution of a conserved passive scalar is described by its mass balance

$$\partial_t r + (\mathbf{u}\cdot\nabla)r - 1/2 n \nabla^2 r = 0 \tag{2.4}$$

with $n$ the diffusion coefficient, the velocity field $\mathbf{u}(\mathbf{x},t)$ of an active medium being determined from its continuum mechanics (2.1), (2.2).

There are also inhomogeneities whose drift motion is not associated with any bulk flow of an active medium. These are disturbances. Heat, an eddy, stress field or discontinuity can move freely in a medium. They are passive in diffusion and active in drift. That enables us to treat them making no reference to a background active medium. In this event, equation (2.4), (2.3) can be retained though with the scalar's drift velocity $\mathbf{v}(\mathbf{x},t)$ entered:

$$\partial_t r + \nabla \cdot (r\mathbf{v}) - 1/2 n \nabla^2 r = 0.$$

Then, on a pattern of continuum mechanics, a dynamic equation can be constructed:

$$r\frac{d\mathbf{v}}{dt}=\tilde{\mathbf{f}}.\tag{2.5}$$

Because of incorporeal character of disturbances (see section 1), the usual way of defining the medium velocity as velocity of its element or corpuscle fails in the case of a delocalized disturbance. Still, by the same reason, disturbances possess the property that enables us to define incremental characteristics as respective weight-averaged quantities. This property is the *self-similarity of a disturbance with respect to scaling its density* i.e. taking $r'(t_1)=kr(t_1)$ entails $r'(t_2)=kr(t_2)$ at a later time $t_2>t_1$. In this event, the force term $\tilde{\mathbf{f}}$ of a dynamic equation can be only a homogeneous first-order function of $r$. Viewing, nevertheless, the disturbance as a body, it can be say, in a more narrow sense that we deal with a gas without hydrostatic pressure. In the next section, continuum mechanics of a density-scaled medium will be build on a regular basis. Among others, the form $\tilde{\mathbf{f}}(r,\mathbf{v})$ of the force term in (2.5) will be specified.

## 3  A route to continuum mechanics

Insofar as ideal fluid is a classical medium, we will seek a motion law of its disturbance (discontinuities including) in the form of classical mechanics. Since it is a turbulent fluid, we will construct a diffusion mechanics. Since the disturbance is smeared or dispersed by the turbulence, it should be a continuum mechanics. Insofar as the system is linear with respect to its density, the technique of Green-function can be conveniently used.

In a quite general approach, temporal evolution of a scalar field $r(\mathbf{x},t)$ can be represented operationally using a transition function $P(\mathbf{x}',t'|\mathbf{x},t)$ of the process, through definition

$$r(\mathbf{x}',t')=\frac{1}{m}\int P(\mathbf{x}',t'|\mathbf{x},t)r(\mathbf{x},t)d\mathbf{x},\tag{3.1}$$

where $m$ is the total "mass" of the scalar

$$m=\int r(\mathbf{x},t)d\mathbf{x}=\text{const}\tag{3.2}$$

and $d\mathbf{x}=dx_1 dx_2 dx_3$. Viewing a conserved scalar as a gas medium, the transition function $P(\mathbf{x}',t'|\mathbf{x},t)$ takes the meaning of the density $r(\mathbf{x}',t')$, the medium has at the time moment $t'$ provided that at the instant $t$ it is wholly concentrated in the point $\mathbf{x}$.

Equation (3.1) is of any value if $P(\mathbf{x}',t'|\mathbf{x},t)$ does not depend on the density $r(\mathbf{x},t)$ of the current state. This is just the property of the scalar field featuring a disturbance. It enables us to give an accurate definition of incremental characteristics. First, the operator $\langle \ | \ $ of averaging over final states is introduced:

$$\langle j | = \frac{1}{m} \int j\, P(\mathbf{x}',t'|\mathbf{x},t) d\mathbf{x}'.$$

Then we define drift velocity

$$\mathbf{v}(\mathbf{x},t) = \lim_{t' \to t} (t'-t)^{-1} \langle \mathbf{x}'-\mathbf{x} |, \qquad (3.3)$$

diffusion coefficient

$$\mathbf{n}_{kl}(\mathbf{x},t) = \lim_{t' \to t} (t'-t)^{-1} \langle (x'_k - x_k)(x'_l - x_l)|, \qquad (3.4)$$

and acceleration

$$\mathbf{a}(\mathbf{x},t) = \lim_{t' \to t} (t'-t)^{-1} \langle \mathbf{v}(\mathbf{x}',t') - \mathbf{v}(\mathbf{x},t) |. \qquad (3.5)$$

Expanding the transition function

$$P(\mathbf{x}'',t''|\mathbf{x},t) = P(\mathbf{x}'',t''|\mathbf{x}',t') + (t-t')\partial_{t'} P(\mathbf{x}'',t''|\mathbf{x}',t')$$
$$+ (x_k - x'_k)\partial_{k'} P(\mathbf{x}'',t''|\mathbf{x}',t')$$
$$+ 1/2 (x_k - x'_k)(x_l - x'_l)\partial_{k'}\partial_{l'} P(\mathbf{x}'',t''|\mathbf{x}',t') + ...$$

and substituting it to the iterated evolution equation (3.1), we come after some manipulations to infinitesimal-incremental form of the mass balance

$$\partial_t r + \nabla \cdot (r\mathbf{v}) - 1/2 \nabla^2 (nr) = 0, \qquad (3.6)$$

where there has been assumed

$$\mathbf{n}_{kl} = n d_{kl}. \qquad (3.7)$$

Expanding velocity

$$\mathbf{v}(\mathbf{x}',t') = \mathbf{v}(\mathbf{x},t) + (t'-t)\partial_t \mathbf{v}(\mathbf{x},t) + (x'_k - x_k)\partial_k \mathbf{v}(\mathbf{x},t)$$
$$+ 1/2 (x'_k - x_k)(x'_l - x_l)\partial_k \partial_l \mathbf{v}(\mathbf{x},t) + ...$$

and substituting it to (3.5), we arrive at local form of the momentum balance

$$\mathbf{a} = \partial_t \mathbf{v} + (\mathbf{v} \cdot \nabla)\mathbf{v} + 1/2 \nabla^2 (n\mathbf{v}) = \mathbf{F}/m, \qquad (3.8)$$

where $\mathbf{F}(\mathbf{x},t)/m$ is the external force mass density.

Equations (3.6), (3.8) represent a model of Brownian diffusion. Indeed, when $\mathbf{F}=0$, $\mathbf{v}=\mathrm{const}$, they reduce to Fourier equation. Distinguishing two kinds of time derivatives makes it possible to construct also a model of superdiffusion.

## 4  Phenomenology of nondifferentiability

Considering a material point of a stochastic medium, its path in the medium may have the shape of a broken line. In that case one must distinguish right-hand $\dot{\mathbf{x}}+$ and left-hand $\dot{\mathbf{x}}-$ time derivatives of the particle's coordinate

$$\dot{\mathbf{x}}+ = \lim_{t' \to t} \frac{\mathbf{x}(t') - \mathbf{x}(t)}{t' - t}, \quad t' > t,$$

$$\dot{\mathbf{x}}- = \lim_{t' \to t} \frac{\mathbf{x}(t) - \mathbf{x}(t')}{t - t'}, \quad t' < t.$$

If the path is broken in each point (so to say, fractal), then the microscopic time derivative can't be determined at all. At last, there may be no a path itself. Anyway, the weight-averaged derivatives (3.3)-(3.5) are ever existent and continuous. But now one must distinguish the forward and backward drift velocities $\mathbf{v}_+$ and $\mathbf{v}_-$, respectively,

$$\mathbf{v}_+(\mathbf{x},t) = \lim_{t'-t \to +0} (t'-t)^{-1} \langle \mathbf{x}' - \mathbf{x} \rangle, \qquad (4.1)$$

$$\mathbf{v}_-(\mathbf{x},t) = \lim_{t'-t \to -0} (t'-t)^{-1} \langle \mathbf{x}' - \mathbf{x} \rangle. \qquad (4.2)$$

where in (4.2) the reverse transition function $P(\mathbf{x}',t'|\mathbf{x},t)$, $t'<t$ is used, which is determined through the respective forward transition function $P(\mathbf{x},t|\mathbf{x}',t')$ by

$$P(\mathbf{x}',t'|\mathbf{x},t) = \frac{P(\mathbf{x},t|\mathbf{x}',t')r(\mathbf{x}',t')}{\int P(\mathbf{x},t|\mathbf{x}',t')r(\mathbf{x}',t')d\mathbf{x}'}.$$

The mass balance (3.6) for the two conjugate flows looks as

$$\partial_t r + \nabla \cdot (r\mathbf{v}_+) - 1/2 \nabla^2 (nr) = 0, \qquad (4.3)$$

$$\partial_t r + \nabla \cdot (r\mathbf{v}_-) + 1/2 \nabla^2 (nr) = 0, \qquad (4.4)$$

respectively, where instead of (3.7)

$$|n_{kl}| = n d_{kl},$$

and

$$n = \text{const}$$

was taken for the case of a homogeneous fluid.

Then, on a pattern of (4.1), (4.2), we define four types of the medium acceleration

$$\mathbf{a}_{++}(\mathbf{x},t) = \lim_{t'-t \to +0} (t'-t)^{-1} \langle \mathbf{v}_+(\mathbf{x}',t') - \mathbf{v}_+(\mathbf{x},t) \rangle,$$

$$\mathbf{a}_{+-}(\mathbf{x},t) = \lim_{t'-t \to -0} (t'-t)^{-1} \langle \mathbf{v}_+(\mathbf{x}',t') - \mathbf{v}_+(\mathbf{x},t) \rangle,$$

$$\mathbf{a}_{-+}(\mathbf{x},t) = \lim_{t'-t \to +0} (t'-t)^{-1} \langle \mathbf{v}_-(\mathbf{x}',t') - \mathbf{v}_-(\mathbf{x},t) \rangle,$$

$$\mathbf{a}_{--}(\mathbf{x},t) = \lim_{t'-t \to -0} (t'-t)^{-1} \langle \mathbf{v}_-(\mathbf{x}',t') - \mathbf{v}_-(\mathbf{x},t) \rangle.$$

Following the derivation of (3.8) that gives four local forms

$$\mathbf{a}_{++} = \partial_t \mathbf{v}_+ + (\mathbf{v}_+ \cdot \nabla) \mathbf{v}_+ + 1/2 \nabla^2 (n\mathbf{v}_+), \qquad (4.5)$$

$$\mathbf{a}_{+-} = \partial_t \mathbf{v}_+ + (\mathbf{v}_- \cdot \nabla) \mathbf{v}_+ - 1/2 \nabla^2 (n\mathbf{v}_+), \qquad (4.6)$$

$$\mathbf{a}_{-+} = \partial_t \mathbf{v}_- + (\mathbf{v}_+ \cdot \nabla) \mathbf{v}_- + 1/2 \nabla^2 (n\mathbf{v}_-), \qquad (4.7)$$

$$\mathbf{a}_{--} = \partial_t \mathbf{v}_- + (\mathbf{v}_- \cdot \nabla) \mathbf{v}_- - 1/2 \nabla^2 (n\mathbf{v}_-). \qquad (4.8)$$

Symmetrizing them, two variants for the dynamic law can be obtained

$$1/2 (\mathbf{a}_{++} + \mathbf{a}_{--}) = \mathbf{F}/m, \qquad (4.9)$$

$$1/2 (\mathbf{a}_{+-} + \mathbf{a}_{-+}) = \mathbf{F}/m. \qquad (4.10)$$

Thus, we arrive at two phenomenological models for transport with diffusion.

In general, we need two velocity functions $\mathbf{v}_+ \neq \mathbf{v}_-$, two kinematic equations (4.3), (4.4) and one dynamic equation (4.9) or (4.10) in order to describe a diffusion process. However, the model (4.9) comprises the forms (4.5), (4.8) with separation of the conjugate flows. Whereas the model (4.10) involves the forms (4.6),(4.7) with entanglement of the flows. Considering the irrotational motion

$$\nabla \times \mathbf{v}_+ = \nabla \times \mathbf{v}_- = 0, \qquad (4.11)$$

we have from (4.5), (4.8)

$$\mathbf{a}_{++} = \mathbf{a}_{--}.$$

In this event, (4.9) degenerates to a single-flow model i.e. there are sufficient for it a one kinematic equation (4.3) or (4.4) plus a dynamic equation given by (4.5) or (4.8), respectively. Still, as before $\mathbf{a}_{+-} \neq \mathbf{a}_{-+}$ and (4.10) represents a double-flow model of diffusion.

## 5 The diffusion force

Formally, the sum of (4.6) and (4.7) in (4.10) can be disentangled if we will pass from $\mathbf{v}_+$, $\mathbf{v}_-$ to $\mathbf{V}$, $\mathbf{w}$ variables, where

$$\mathbf{V} = 1/2\left(\mathbf{v}_+ + \mathbf{v}_-\right),$$
$$\mathbf{w} = 1/2\left(\mathbf{v}_- - \mathbf{v}_+\right).$$

Those velocities receive their meaning from the kinematical part of the problem. Adding conjugate Fokker-Planck equations (4.3) and (4.4) we obtain a continuity equation

$$\partial_t r + \nabla \cdot (r\mathbf{V}) = 0 \qquad (5.1)$$

with the median $\mathbf{V}$ taking the sense of the complete velocity. Subtracting (4.3) from (4.4) and integrating it over $\mathbf{x}$, we come to Fick's relation

$$r\mathbf{w} = -1/2n\nabla r \qquad (5.2)$$

with the "saltus" $\mathbf{w}$ receiving the sense of the diffusion velocity, where $r \to 0$, $\nabla r \to 0$ when $\mathbf{x} \to \pm\infty$ was used. So, $\mathbf{v}_+$ acquires the meaning of regular component of the complete velocity:

$$\mathbf{V} = \mathbf{v}_+ + \mathbf{w}.$$

From (4.5), (4.8) we get for (4.9) in terms of $\mathbf{V}$, $\mathbf{w}$:

$$\partial_t \mathbf{V} + (\mathbf{V} \cdot \nabla)\mathbf{V} = \mathbf{F}/m - (\mathbf{w} \cdot \nabla)\mathbf{w} + 1/2n\nabla^2 \mathbf{w}. \qquad (5.3)$$

From (4.6), (4.7) we get for (4.10):
$$\partial_t \mathbf{V} + (\mathbf{V}\cdot\nabla)\mathbf{V} = \mathbf{F}/m + (\mathbf{w}\cdot\nabla)\mathbf{w} - 1/2 n\nabla^2 \mathbf{w}. \qquad (5.4)$$

When expressed in $\mathbf{V}$, $\mathbf{w}$ variables, the diffusion factor is not present explicitly in the kinematical equation (5.1). In return, an addition $\pm\mathbf{F}'$ to the external force $\mathbf{F}$ appears in the dynamic equations (5.3) and (5.4), which depends on the diffusion velocity:
$$\mathbf{F}' = 1/2 m\nabla\left(\mathbf{w}^2 - n\nabla\cdot\mathbf{w}\right), \qquad (5.5)$$

where $\nabla\times\mathbf{w}\equiv 0$ was used. In those terms we have for (5.3)
$$m\frac{d\mathbf{V}}{dt} = \mathbf{F} - \mathbf{F}' \qquad (5.6)$$

and for (5.4)
$$m\frac{d\mathbf{V}}{dt} = \mathbf{F} + \mathbf{F}'. \qquad (5.7)$$

Now, both of the models are represented uniformly: the single-flow model (4.9) – by the set of equations (5.1), (5.6) with (5.5), (5.2), the composite-flow model (4.10) – by the set (5.1), (5.7) with (5.5), (5.2). The kinematical diffusion factor is retained in the initial values of the problem:
$$\begin{aligned} r(\mathbf{x},0) &= r_0(\mathbf{x}), \\ \mathbf{V}(\mathbf{x},0) &= \mathbf{v}_{+0}(\mathbf{x}) - 1/2 n\nabla\ln r_0(\mathbf{x}), \end{aligned} \qquad (5.8)$$

where
$$\mathbf{v}_{+0}(\mathbf{x}) = \mathbf{v}_{+}(\mathbf{x},0).$$

Assuming in a one-dimensional case
$$r \propto \exp\left[-a(x-x_0)^2\right], \qquad a>0,$$

the mutual direction of the diffusion force (5.5) and the diffusion flow (5.2) is specified by the sign of the integral:
$$\frac{1}{m}\int r\mathbf{w}\mathbf{F}'d\mathbf{x} = na\int r\mathbf{w}^2 dx \geq 0.$$

So, in (5.6) the diffusion force tends to oppose the diffusion smearing of the distribution (initiated by the kinematical diffusion factor (5.2) through (5.8)). In (5.7) the diffusion force promotes the kinematical smearing of the distribution. This recommends to try (5.7) as a model of the so-called hyperdiffusion observed in experiments with ordinary fluids.

The common feature of all diffusion processes is the Fick's relation (5.2). It implies the following correlations

$$\mathrm{cov}(x_k, w_l) = \boldsymbol{d}_{kl} \boldsymbol{n}/2,$$

$$\mathrm{cov}(x_1, F_1) = -m \langle w_1^2 \rangle,$$

where

$$\mathrm{cov}(a,b) = \frac{1}{m} \int (a - \langle a \rangle)(b - \langle b \rangle) \boldsymbol{r}(\mathbf{x}) d\mathbf{x},$$

$$\langle c \rangle = \frac{1}{m} \int c \boldsymbol{r}(\mathbf{x}) d\mathbf{x}.$$

With this, Cauchy-Bunyakovsky inequality

$$\mathrm{cov}(a,a)\mathrm{cov}(b,b) \geq \mathrm{cov}^2(a,b)$$

provides the indeterminacy relation

$$\langle (x_k - \langle x_k \rangle)^2 \rangle \langle (w_l - \langle w_l \rangle)^2 \rangle \geq \boldsymbol{d}_{kl}(\boldsymbol{n}/2)^2$$

and inequality

$$\langle (F_1' - \langle F_1' \rangle)^2 \rangle \langle (x_1 - \langle x_1 \rangle)^2 \rangle \geq m^2 \langle w_1^2 \rangle^2.$$

Combining these yields

$$\langle (F_1')^2 \rangle^{1/2} \geq m(\boldsymbol{n}/2)^2 / \langle (x_1 - \langle x_1 \rangle)^2 \rangle^{3/2}, \qquad (5.9)$$

where $\langle \mathbf{w} \rangle = 0$, $\langle \mathbf{F}' \rangle = 0$ were used. Inequality (5.9) indicates that high local concentrations of a scalar favor a hyperdiffusion. This agrees with the experimental observation [3].

## 6 The wave-function representation

The Fick's law (5.2) provides a potential for the diffusion velocity:

$$\mathbf{w} = -\boldsymbol{n} \nabla \ln \sqrt{\boldsymbol{r}}.$$

Considering irrotational motion (4.11), a potential A for **V** can be defined

Also we assume

$$\mathbf{V} = n\nabla A.$$

$$\mathbf{F} = -\nabla U.$$

Under the substitution

$$\Psi = \sqrt{r}\exp(iA)$$

the set (5.1), (5.2), (5.4) of nonlinear equations rolls up into a complex-valued linear equation

$$-1/2\,n^2\nabla^2\Psi + (U/m)\Psi = in\partial_t\Psi. \qquad (6.1)$$

This is the Cauchy-Lagrange integral of the continuum mechanics. It has the form of the Schroedinger equation. Simultaneously, the evolution equation (3.1) turns into the amplitude form:

$$\Psi(\mathbf{x}',t') = \int K(\mathbf{x}',t'|\mathbf{x},t)\Psi(\mathbf{x},t)d\mathbf{x},$$

where, unlike $P$, the kernel $K$ does not depend on the (instantaneous) memory $\mathbf{v}$; that is, $K$ has the meaning of an evolution law.

A single flow model (5.1), (5.2), (5.3) convolves into a real-valued linear equation

$$1/2\,n^2\nabla^2\Phi + (U/m)\Phi = 1n\partial_t\Phi$$

under the substitution

$$\Phi = \sqrt{r}\exp(1A).$$

## 7 Diffusion kinetics

In this section explicit expressions for $r(\mathbf{x},t)$ and $\mathbf{V}(\mathbf{x},t)$ are given. That enables us to compare mathematical models (4.9), (4.10) with macroscopic experiments on diffusion of a disturbance.

The following particular solutions can be obtained. Resolving equations (5.1), (5.2), (5.3) we get for one-dimensional case of the single-flow:

$$r(x,t) = \frac{1}{\sqrt{2ps_0^2(1 + nt/s_0^2)}}\exp\left[-\frac{(x - u_+ t)^2}{2s_0^2(1 + nt/s_0^2)}\right], \qquad (7.1)$$

$$V(x,t) = u_+ + n\frac{x - u_+ t}{2s_0^2(1 + nt/s_0^2)}.$$

Resolving equations (5.1), (5.2), (5.4) we have for one dimensional case of the composite-flow:

$$r(x,t) = \frac{1}{\sqrt{2ps_0^2\left(1+\left(\frac{nt}{2s_0^2}\right)^2\right)}} \exp\left[-\frac{(x-u_0 t)^2}{2s_0^2\left(1+\left(\frac{nt}{2s_0^2}\right)^2\right)}\right], \quad (7.2)$$

$$V(x,t) = \left[u_0 + \left(\frac{n}{2s_0^2}\right)^2 xt\right] / \left[1+\left(\frac{n}{2s_0^2}\right)^2 t^2\right]$$

In order to characterize diffusion kinetics of a scalar on the whole, the square broadening of its distribution is defined:

$$s^2 = \left\langle (x-\langle x \rangle)^2 \right\rangle = \frac{1}{m}\int (x-\langle x \rangle)^2 r(x,t)dx,$$

where *m* is the total mass (3.2) of the scalar and

$$\langle x \rangle = \frac{1}{m}\int x r(x,t) dx.$$

So, we have for the single-flow model (7.1):

$$s = \sqrt{s_0^2 + nt}. \quad (7.3)$$

This is peculiar to Brownian kinetics of diffusion. The composite-flow model (7.2) gives

$$s = \sqrt{s_0^2 + \left(\frac{nt}{2s_0^2}\right)^2}. \quad (7.4)$$

The latter is peculiar to hyperdiffusion.

## 8  Matching with the experiment

There are two types of diffusion kinetics observed in experiments on evolution of a disturbance in a stochastic environs. That enables us to discriminate two variants (4.9) and (4.10) of the mathematical model. Macroscopic realization of the dispersion of a point defect gives us also occasion to verify our physical model of a quantum particle.

Consider experiments [3] on evolution of a heat pulse in turbulent jets. At the outset, we have an instantaneous point source of heat, which is imbedded in a turbulent fluid. Later on, the narrow distribution of temperature is smeared up. A linear growth of the half-width $s$ of the distribution with time

$$s \propto nt \qquad (8.1)$$

is observed in the initial stage. Afterwards, it follows the Brownian kinetics

$$s \propto \sqrt{nt} \qquad (8.2)$$

These can be interpreted as related with two forms of a disturbance. Brownian kinetics (8.2) corresponds to usual heat conductivity. This is peculiar to the medium, which consists of corpuscles. Comparing (8.2) with (7.3) for $s_0 \ll nt$ we see that it is described by the single-flow model (4.9). The regime of hyperdiffusion (8.1) has the kinetics of convection. This can be explained in terms of a heat soliton, which is split and dispersed in the turbulent continuum. Comparing (8.1) with (7.4) for $s_0 \ll nt$ we see that the composite-flow model (4.10) is suitable here.

Another macroscopic realization is given by the dislocation plasma in metals. In [4] it has been described by the Schroedinger continuum mechanics (6.1).

## 9  Distribution of a singularity

Now, both the mechanism of hyperdiffusion and mechanics of a quantum particle can be elucidated. We consider a singular disturbance or a discontinuity of a mechanical medium (of a substratum). It is supposed to be split and dispersed under the action of turbulence forming the plasma or dispersion of the discontinuity. This dispersion will be further referred to as a discontinuum. We suggest that the discontinuum evolves according to the double-flow model (5.4) of diffusion mechanics. You see from (7.4) that it spreads in the space convectively with the average speed of the spreading evaluated as

$$c^* = \partial_t s \to n/s_0. \qquad (9.1)$$

Physically, this is none other than the wave of plastic deformation of the substratum. In the spirit of our approach, the motion of the discontinuum in the substratum will be viewed as acoustic wave in the very discontinuum. (Compare this with the discussion in [5] where the plastic wave in solids has been interpreted as a perturbation wave in a "dislocation gas".) Indeed, as is known, the acoustic wave in a gas is accompanied by transferring some amount of the medium material. The more it is true in respect of the discontinuity plasma. The discontinuum altogether serves as a medium to support a wave and also it is spread itself over the substratum due to this wave.

Following [6], we take a portion of the discontinuum formed from a point defect and treat as an elastic body. From the dynamic equation (5.4), (5.2) the tensor of diffusion stresses of the discontinuum can be extracted:

$$p_{ij} = (n/2)^2 r \partial_i \left( \partial_j r / r \right). \tag{9.2}$$

Insofar as the defect's density $r$ has the meaning of volume, for a free defect it is factorized:

$$r(\mathbf{x}) = r_1(x_1) r_2(x_2) r_3(x_3).$$

Hence, the nondiagonal elements of the tensor (9.2) are vanishing. The latter means that the discontinuum supports only longitudinal waves. The wave of the point discontinuum models the de Broglie wave of a particle. Otherwise stating, *it is the wave of plastic deformation of a substratum that serves as a mechanical analogy of the de Broglie wave.*

Now, we may deduce the relation between the wave length $l$ and the velocity $u$ of translational motion of the discontinuum in purely mechanical terms. With this end, we assume that the energy of motion $1/2 m u^2 a$ of the discontinuum is equal to its vibration energy $1/2 m (c'/l)^2 A^2$, where $m$ is the total mass of the discontinuum, $c'$ the speed of the sound wave in the discontinuum and $A$ the amplitude of its vibration. That gives

$$l = c' A / u.$$

The definition $c'_i = \left( -\partial p_i / \partial r \right)^{1/2}$ with (9.2) and (7.2) returns us to (9.1), though

$$c' = 1/2 n \sqrt{-\partial^2 \ln r} = 1/2 n / s.$$

Whence, supposing that

$$A \sim s,$$

we get the de Broglie relation

$$l \sim n / u.$$

## 10  To complete the discontinuity model of a quantum particle

Let $\Lambda$ is the size of a drop of a material impurity dissolved in a fluid or the size of a discontinuity of a substratum; $l$ is the size of a molecule of the impurity or the size of a splinter of the discontinuity. The diffusion coefficient of an impurity is known to depend on the size $l$ of its molecule. The discontinuity plasma is taken as a mechanical model for a quantum particle. To be a consistent model, the diffusion coefficient $n$ of the discontinuity plasma must depend on $\Lambda$, not on $l$. Indeed, the respective coefficient in quantum mechanics is dependent on the total mass $m$ of the particle:

$$n = \hbar / m. \tag{10.1}$$

So, we deal here with a special kind of discontinuity. These are the *cavitons* – dilatational inclusions of a void or of the quiescent fluid in a turbulent fluid associated with centers of turbulence perturbation. Splitting of a caviton implies a change in the phase state within the core of the caviton, the volume of the core remaining invariant [7]. Consequently, we have for cavitons

$$l = \Lambda.$$

It can be shown that the distribution of a caviton can always be represented as the sum of the splinters, each of which reproduces the structure of the original caviton. Let $\Lambda$ be the core's volume. Then the mass of a localized caviton is equal to $\nu\Lambda$, where $\nu$ is the fluid density. That provides prerequisites for deriving formula (10.1). Otherwise, a distributed caviton behaves itself as a consistent integrity and should not be explained in terms of a discrete model.

Another question of the model is that of the "wave-function collapse". Consider a void split and distributed over a fluid. Let in some place of the fluid there is introduced a quantity of the energy. That leads to a local drop in the fluid pressure and hence to recollection of the entire void in this place – at the expense of shrinking down all other fragments of the void. The speed of the process is comparable with the speed of a compression wave in an incompressible fluid i.e. it tends to infinity. A lot of intriguing details relevant to this phenomenon is observed in experiments [8] with vortex turbulence of ordinary liquids. Such is a continuum mechanics model for the collapse of the wave function caused by intervening of some external factor.

PR/E-email
15 July 1998  22:37
dmitr@cc.nifhi.ac.ru
pre@aps.org
RE:  ES6368

ES6368
Towards a mechanical analogy of a quantum particle: Turbulent advection of a
Dmitriyev,Valery P./

Dear Dr. Dmitriyev:

The above manuscript has been reviewed by one of our
referees. Comments from the report are enclosed.

These comments suggest that the present manuscript
is not suitable for publication in the Physical Review.

Yours sincerely,
David A. Weitz
Associate Editor
Physical Review E
================================================================

REPORT

This paper presents interesting new results about an
analogy between a very peculiar case of fluid mechanics
and quantum mechanics using a stochastic approach and,
per se, would be worth publishing.

Yet, in the present form, the general level of
presentation is not high and, more important, the
physical examples are not developed enough: the result is
that the paper appears more as a mathematical exercise
than a new deep piece of physics.

My suggestion is that the author improves the physical
aspects and submits a revised version for further
examination.
================================================================
rvw_stndrds_au_pre.asc
revised 3/98                                    PHYSICAL REVIEW E

The earlier version of this work has been delivered in 1996 to the British Society for Philosophy of Science Meeting "Physical Interpretation of Relativity Theory".